\documentclass[journal = jpccck,manuscript=article]{achemso}
\usepackage{amsmath}

\title[]{DFT Studies of Adsorption of Benzoic Acid on the Rutile $(110)$ Surface: Modes and Patterns}
\author{Xiang Zhao (Ivan)}
\email{ucapxzh@ucl.ac.uk}
\affiliation[LCN]{London Centre for Nanotechnology, 17-19 Gordon Street, London, WC1H 0AH}
\alsoaffiliation[UCL]{Department of Physics \& Astronomy, University College London, Gower Street, London, WC1E 6BT}
\author{David R. Bowler}
\affiliation[LCN]{London Centre for Nanotechnology, 17-19 Gordon Street, London, WC1H 0AH}
\alsoaffiliation[UCL]{Department of Physics \& Astronomy, University College London, Gower Street, London, WC1E 6BT}
\alsoaffiliation[MANA Satellite, UCL]{UCL Satellite, International Centre for Materials Nanoarchitectonics (MANA), National Institute for Materials Science (NIMS),
1-1 Namiki, Tsukuba, Ibaraki 305-0044, Japan}
\email{david.bowler@ucl.ac.uk}

\begin{document}

\section{Abstract}
Adsorption of benzoic acid on the $(110)$ surface of rutile, both unreconstructed and $(1 \times 2)$-reconstructed ones, at saturation coverage of one benzoate per two adjacent five-coordinated Ti$_{5\text{c}}$ sites has been studied using DFT simulations, as implemented through the Vienna Ab initio Simulations Package (VASP). In order to study the effects of hydrogen bonding and Van der Waals forces in influencing the relative stabilities of different adsorbate overlayer superstructures, these studies were performed through Local Density Approximation (LDA), Generalized Gradient Approximation(GGA) and DFT-D2. Through the calculations, it was found out that although the optimized structures corresponded with the proposed models for the experimental results, the relative energetic stabilities of different overlayer structures have shown some differences with the experimental results. In the GGA calculations, the overlayer structures involving the benzene rings aligned along the $(001)$-direction were shown to be more stable than those aligned perpendicular to it, regardless of whether the benzoates were arranged in $(1 \times 1)$ or $(1 \times 2)$ symmetries on the $(1 \times 1)$-unreconstructed surface, or on the $(1 \times 2)$-reconstructed surface. These are in contrast to the experimental studies using STM, whose observations revealed that configurations with the benzene rings aligned along the $(1\bar{1}0)$-direction when adsorbed on the $(1 \times 2)$-reconstructed surface.

\section{Introduction}
Titanium dioxide (TiO$_{2}$) has a history of applications since the early 19th century. Although its initial usage was confined to use as a white pigment, its applications diversified over the course of the last few decades, including but not limited to biological implants, photocatalysis and dye-sensitized solar cells (DSSCs), based on the refractive, biocompatible, photocatalytic and photoelectric properties of the material \cite{Diebold}. Recent emphasis on the latter applications spurred research work in the surface properties and reactions of crystalline TiO$_{2}$, understanding of which is crucial in their realization.

In the case of DSSCs, this often involves the interactions of these surfaces with sensitizing dyes, such as triscarboxy-ruthenium terpyridine [Ru(4,4',4''-(COOH)$_{3}$-terpy)(NCS)$_{3}$], the "black dye". Recent developments have also been made in bio-sensitized solar cells (BSSCs)\cite{BSSC}, which are similar to DSSCs in operating principles, but with the synthetic dye often used in the latter replaced by one that is biologically derived. One class of such biomolecules are the cyclic tetrapyrroles, which encompass porphyrins and chlorins. Many of such compounds contain a ring structure, with one or more carboxylic functional groups which play importance roles in anchoring the dye to the TiO$_{2}$ surfaces, benzoic acid's adsorption on TiO$_{2}$ can thus offer an exemplar case study for how this entire class of compounds can interact with TiO$_{2}$ surfaces. 

Out of all the TiO$_{2}$ surfaces, the rutile $(110)$ is the most energetically stable and naturally occurring, and thus forms the substrate surface of choice in such adsorption studies. The rutile $(110)$ surface, though not prone to reconstruction, can undergo $(1 \times 2)$ reconstruction upon annealing under UHV conditions, with the models proposed including added Ti$_{2}$O$_{3}$ and Ti$_{3}$O$_{6}$ rows. As such, the focus of our studies shall be on the adsorption of benzoic acid on these two types of rutile $(110)$ surfaces, in terms of the binding mode and the overlayer patterns.

Experimental studies have been performed on such adsorptions, using LEED, ESDIAD and STM by Guo et al \cite{BenzRutSTM}, and later by Grinter et al using STM \cite{BenzRutNew}. In Guo's studies, the optimal adsorption mode observed was in line with those of other small carboxylates i.e. dissociative bidentate bridging binding (BB), involving the two carboxyl O atoms, after deprotonation on one of them, binding to two neighbouring five-coordinated Ti sites (Ti$_{5\text{c}}$) along the $[001]$ rows of such sites. 

In terms of adsorption patterns, through LEED\cite{BenzRutSTM}, it was also observed that a p$(2 \times 1)$ pattern was formed. When studied using STM however, a different picture emerged, as a p$(2 \times 2)$ pattern was observed instead. Such discrepancy could be accounted for by the fact that LEED is a scattering technique dependent upon the charge distribution of the surface, while STM surveys the apex of the surface and the adsorbates, and the fact that the phenyl rings of the benzoates were rotated to align with the $[0\bar{1}1]$ direction to form dimer structures. The rotation of the phenyl rings and the resultant dimer structures are due to the interactions of the lower phenyl H atoms and the $\pi$ orbital of the neighbouring benzoate, facilitated by the bridging O$_{2\text{c}}$ anions.

Building upon these findings, a later STM study of these adsorptions was performed\cite{BenzRutNew}, with the aims of studying orientations of the benzene rings of the adsorbate overlayer, as well as the adsorption of benzoates on the $(110)$-$(1 \times 2)$ reconstructed surface. The binding geometry of the acid was once again confirmed, but the elongation of the phenyl rings along the $[1\bar{1}0]$ direction was only significant in the case of the reconstructed surface, due to improved hydrogen bonding of the phenyl H atoms to the surface O atoms as a result of closer proximity. There was also no formation of dimer structures at high coverages as reported in the much earlier study - which could be due to the higher dosing temperature used this time, nor was there any observation of adsorption of dissociated H$^{+}$ ions. On the $(110)$-$(1 \times 2)$ reconstructed surface, it was also observed that benzoic acid adsorbs and binds the same way as on the unreconstructed surface.

Although DFT studies of the adsorption of benzoic acid on the rutile $(110)$ surface had been performed by Troisi et al \cite{BenzTiO2DFT}, these were performed with the aim of comparing the effects of using different DFT implementations on the computed results on electronic structures of such benzoate-TiO$_{2}$ complexes, rather than the physical structure of the adsorbates on the surface which forms the main theme of experimental research work. In our work, we will expand upon this topic by investigating the proposed models for surface adsorbate structures and patterns described in the experimental work, using DFT as implemented through VASP. After presenting our methodology, we calculate and compare the energetics of these configurations obtained using LDA, GGA and vdW, as well as producing high resolution simulated STM images to provide comparisons with these results, before concluding.

\section{Methodology}

Our DFT investigations were carried out using the Vienna Ab initio Simulations Package (VASP)\cite{VASP}\cite{VASP2}. Out of all the elemental pseudopotential files, oxygen has the highest $E_{\text{cut}}$ value at 395.700~eV, hence this value was chosen as the cut-off energy $E_{\text{cut}}$ in all our simulations. Iterative electronic relaxations are done such that energy differences ($E_{\text{diff}}$) between two successive steps should not exceed $10^{-6}$~eV, while relaxations are said to be achieved when the RMS force on each atom falls below $0.03$~eV/\AA. 

In our calculations, LDA, GGA(PBE) and GGA+DFT-D2-based methods were used to give three different sets of results for comparison of the effects of hydrogen bonding and Van der Waals forces, as GGA and DFT-D2 methods can account for these forces much better, as the interactions as a result of these forces were implied in the experimental STM studies \cite{BenzRutNew}.

In order to simulate the rutile TiO$_{2}$ substrate, an 8 Ti layer slab terminated by two $(110)$ surfaces on each end was chosen, as studies of the rutile $(110)$ surface through DFT have revealed that surface energies converge at around this thickness \cite{TiO2110STMDFT}. This was done after performing bulk relaxation of the rutile crystal, where K-sampling values of $12 \times 12 \times 16$ were used, in order to reflect the approximate inverse ratio of the dimensions of the real space unit cell. The optimized lattice constants obtained through LDA calculations were found to be  $a = 4.546 \text{~\AA}$ and $c = 2.925 \text{~\AA}$ for rutile, reflecting a contractions of $0.03$~\AA\ to $0.05$~\AA\ from the experimental values, which are expected for DFT using LDA as reported in other research literature, as well as being within $\pm 2 \%$ error margin \cite{RutSurfOptim}. As for GGA based calculations, the optimized lattice constants were $a = 4.615 \text{~\AA}$ and $c = 2.966 \text{~\AA}$, representing an expansion of 0.04~\AA, which is expected due to GGA's underestimation of cohesive energies of insulators.

The new lattice parameters were then used to set up the slabs for surface relaxations. 8 Ti layer slabs were set up using four such cells with the terminated ends autocompensated. For the $(1 \times 2)$-reconstructed surface, we used the Ti$_{3}$O$_{6}$ added-row structure as assumed in \cite{BenzRutNew}. We then set the benzoates up such that they correspond to the following six configurations, which involve the phenyl rings being aligned either with the $[001]$ direction(denoted $_{\parallel}$) or the $[1\bar{1}0]$ direction (denoted $_{\perp}$), as well as being aligned such that they form $(1 \times 1)$-periodic patterns (denoted $_{\text{i.p.}}$), or with the neighbouring $[001]$-rows of benzoates aligned one Ti$_{5\text{c}}$ step out with respect to each other (denoted $_{\text{o.p.}}$).

The different monolayer patterns are denoted as follows, with the relaxed GGA structures shown in the relevant figures (LDA and GGA relaxed structures show little different):
\begin{itemize}
	\item BB$_{\parallel, \text{ i.p.}}$: Phenyl rings of the benzoates are aligned with the $[001]$ direction, with the benzoates of neighbouring $[001]$ rows in step with each other, as seen in Figure \ref{ParaInGGA}
	\item BB$_{\parallel, \text{ o.p.}}$: Phenyl rings of the benzoates are aligned with the $[001]$ direction, with the benzoates of neighbouring $[001]$ one Ti$_{5\text{c}}$ site out of step with each other, as seen in Figure \ref{PerpInGGA}
	\item BB$_{\perp, \text{ i.p.}}$: Phenyl rings of the benzoates are perpendicular to the $[001]$ direction, with the benzoates of neighbouring $[001]$ rows in step with each other, as seen in Figure \ref{ParaOutGGA}
	\item BB$_{\perp, \text{ o.p.}}$:  Phenyl rings of the benzoates are perpendicular to the $[001]$ direction, with the benzoates of neighbouring $[001]$ one Ti$_{5\text{c}}$ site out of step with each other, as seen in Figure \ref{PerpOutGGA}
	\item BB$_{\parallel, \, 1\times2}$: Phenyl rings of the benzoates are aligned with the $[001]$ direction, using the $(1 \times 2)$-reconstructed $(110)$ surface instead, as seen in Figure \ref{ParaReconGGA}
	\item BB$_{\perp, \, 1\times2}$: Phenyl rings of the benzoates are perpendicular to the $[001]$ direction, using the $(1 \times 2)$-reconstructed $(110)$ surface instead, as seen in Figure \ref{PerpReconGGA}
\end{itemize}

These collections of atoms are padded by ~15~\AA\ of vacuum between the topmost and the bottom-most atoms of the vertically adjacent simulation cells, in order to ensure lack of interaction between the atoms of these cells. To calculate the adsorption energies ($E_{\text{ads}}$), the energies of three simulation cells of the above defined dimensions, consisting of the slab and the adsorbate ($E_{\text{slab}+\text{mol}}$), just the slab ($E_{\text{slab}}$) and just the adsorbate ($E_{\text{mol}}$) in neutral form respectively, were first calculated, obtaining $E_{\text{ads}}$ by:

\begin{eqnarray}
E_{\text{ads}} = E_{\text{mol + slab}} - (E_{\text{mol}} +  E_{\text{slab}})  \label{Eads}
\end{eqnarray} 

As the latest STM studies \cite{BenzRutNew} did not report H adsorption on the protruding bridging O$_{2\text{c}}$s, these simulations were also run with the dissociated H removed, with stabilities of each mode compared in terms of the total energy of the simulation cell $E_{\text{cell}}$. In addition, the ionic adsorption energies of $E_{\text{ion, ads}}$ the different modes will be compared against one another, in a similar fashion to \eqref{Eads}, with $E_{\text{mol}} $ being replaced by that corresponding to the energy of a single benzoate anion in the simulation cell. This gives us completeness in the comparisons for adsorption energies.

Upon relaxations of these different modes of adsorptions under the above described set ups, simulated STM images were then generated based on the electronic structures of the relaxed adsorbate-surface complex, at a bias voltage of +1.5V.

\section{Results}

\subsection{LDA Calculations}

For the adsorption structures, the optimized structures for the six BB modes corresponded largely with the models proposed for the STM images \cite{BenzRutNew}, in that the carboxylate group remained anchored via two Ti$_{5\text{c}}$-O$_{\text{carboxy}}$ bonds, while the benzene rings remained aligned more or less along the $[001]$ and the $[1\bar{1}0]$ directions. Slight deviations were observed in the BB$_{\parallel, \text{i.p.}}$ and the BB$_{\parallel, \, 1\times2}$ modes, in that the benzene rings were not exactly aligned along the $[001]$ directions and instead being slightly rotated about the $[110]$ axis. In addition to the slight rotations, in all of the BB$_{\parallel}$ modes, the benzene rings underwent slight planar rotations about the $[1\bar{1}0]$ axis. These are likely due to the sideways repulsions between the H atoms on the benzene rings.

In terms of adsorption energetics, the adsorption energies ($E_{\text{ads}}$) were obtained (with the values obtained through GGA calculations) as displayed in Table \ref{DissAds}. When the dissociated Hs were removed, ionic adsorption energies ($E_{\text{ion, ads}}$) were obtained and presented in Table \ref{IonAds}.

\begin{table} 
\begin{tabular}{|p{3.5cm}|p{3.5cm}|p{3.5cm}|p{3.5cm}|}  
\hline
\textbf{Adsorption Mode} & E$_{\text{ads}}$(\textbf{dis}) & E$_{\text{ads}}$(\textbf{no H})  \\
\hline
BB$_{\parallel, \text{ i.p.}}$ (Figure \ref{ParaInGGA}) & -2.30 & -3.93 \\
\hline
BB$_{\perp, \text{ i.p.}}$  (Figure \ref{PerpInGGA}) & -3.49 & -3.99 \\
\hline
BB$_{\parallel, \text{ o.p.}}$ (Figure \ref{ParaOutGGA}) & -0.31 & -3.01 \\
\hline
BB$_{\perp, \text{ o.p.}}$ (Figure \ref{PerpOutGGA}) & -1.25 & -4.23 \\
\hline
BB$_{\parallel, \, 1\times2}$ (Figure \ref{ParaReconGGA}) & -2.20 & -1.94 \\
\hline
BB$_{\perp, \, 1\times2}$ (Figure \ref{PerpReconGGA}) & -2.00 & -0.91 \\
\hline
\end{tabular} 
\caption{Adsorption energies for dissociative adsorptions of benzoic acid on the rutile $(110)$ surface, in~eV/\AA, as calculated through LDA, with and without the co-adsorbed H. \label{DissAds}}
\end{table}

From the figures for adsorption energetics alone, the LDA results did not completely corroborate with the STM studies \cite{BenzRutSTM,BenzRutNew}. In the former, it was observed that the BB$_{\parallel, \text{ i.p.}}$ mode should be more stable than the BB$_{\perp, \text{ i.p.}}$ configuration, our current results of DFT calculations however contradict these. This could possibly be explained by the fact in such an arrangement, the benzene rings are close to each other, resulting in repulsion between the neighbouring rings, thereby destablizing the configuration. By rotating, this source of destablization was removed. 

In the BB$_{\text{o.p.}}$ superstructures, the results showed agreement with the observations that the benzene rings would be rotated along the $[1\bar{1}0]$ direction, with  $E_{\text{ads}}$ (BB$_{\perp, \text{ o.p.}}$) being 0.06~eV more stable than the value for $E_{\text{ads}}$ (BB$_{\parallel, \text{ o.p.}}$). The difference however, was much smaller than expected taking into consideration of the predomination of the BB$_{\perp, \text{ o.p.}}$ pattern in the STM images.

For adsorption on the reconstructed surface, stable dissociative bidentate bridging adsorption of benzoic acid on the surface was observed at the Ti$_{5\text{c}}$ sites between the reconstructed ridges along the $[001]$ directions. The adsorption energy values of the BB$_{\parallel, \, 1\times2}$ and the BB$_{\perp, \, 1\times2}$ patterns however, still contradicted the  STM observations, with the non-rotated modes being almost ~0.5~eV more stable than the rotated ones.

When the dissociated hydrogens were removed, ionic adsorption energies ($E_{\text{ion, ads}}$) were obtained using the method described in \eqref{Eads} and presented in Table \ref{IonAds}.

\begin{table} 
\begin{tabular}{|p{3.5cm}|p{3.5cm}|p{3.5cm}|p{3.5cm}|}  
\hline
\textbf{Adsorption Mode} & E$_{\text{ads}}$(\textbf{dis}) & E$_{\text{ads}}$(\textbf{no H}) \\
\hline
BB$_{\parallel, \text{ i.p.}}$ (Figure \ref{ParaInGGA})  & -2.78 & -3.16 \\
\hline
BB$_{\perp, \text{ i.p.}}$ (Figure \ref{PerpInGGA}) & -1.29 & -1.55 \\
\hline
BB$_{\parallel, \text{ o.p.}}$ (Figure \ref{ParaOutGGA}) & -1.56 & -1.10 \\
\hline
BB$_{\perp, \text{ o.p.}}$ (Figure \ref{PerpOutGGA}) & -1.31 & -0.65 \\
\hline
BB$_{\parallel, \, 1\times2}$ (Figure \ref{ParaReconGGA}) & -1.15 & -1.80 \\
\hline
BB$_{\perp, \, 1\times2}$ (Figure \ref{PerpReconGGA}) & -0.70 & -1.25 \\
\hline
\end{tabular} 
\caption{Adsorption energies for adsorptions of benzoates on the rutile $(110)$ surface, in~eV/\AA, as calculated through GGA, with and without the co-adsorbed H. \label{IonAds}}
\end{table}

When the same simulations were rerun with the dissociated hydrogens removed, however, slightly different pictures emerged as the BB$_{\perp, \, 1\times2}$ mode became more energetically stable than the BB$_{\parallel, \, 1\times2}$ mode, as was observed in the STM studies. The BB$_{\text{i.p.}}$ modes also became energetically comparable, while the BB$_{\parallel, \text{o.p.}}$ mode became slightly more energetically favourable than the BB$_{\perp, \text{o.p.}}$. This suggests that the presence and absence of H atoms has influences on the overall stabilities of the adsorption structures,even without hydrogen bonding taken into account.

The discrepancies between the calculated results and the experimental findings can be attributed to LDA as a method for DFT calculations that does not include hydrogen bonding, as well as $\pi$-$\pi$ interactions, which in this case are those between neighbouring benzene rings. These are interactions that were proposed as significant factors in stablizing the adsorption structures, thougheven without being taken into account, rotation of the benzene rings was seen to have major effects on the energetic stabilities of the adsorptions.

\subsection{GGA Calculations}
\begin{figure}[!hbtp]
\centering
\includegraphics[width = 0.30\textwidth]{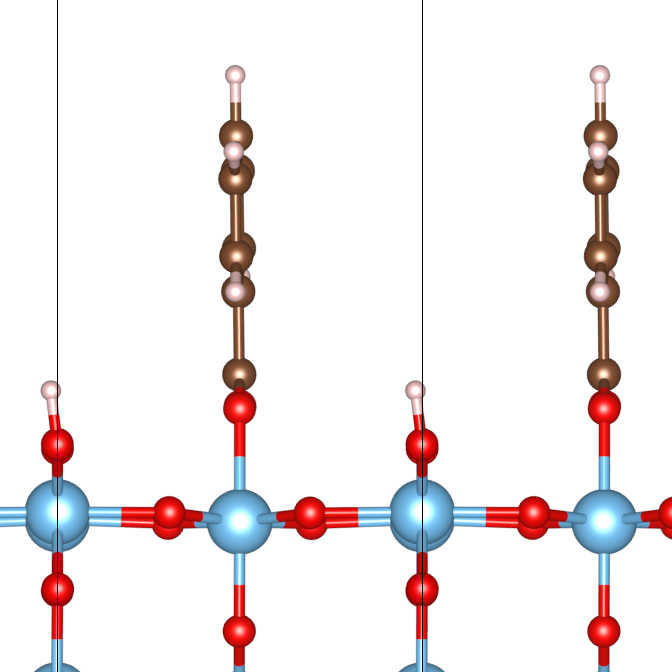}
\includegraphics[width = 0.30\textwidth]{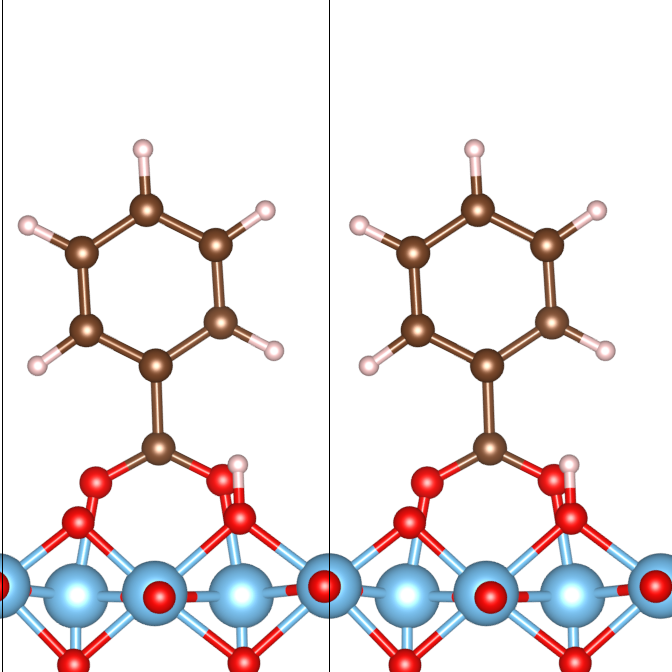}
\includegraphics[width = 0.30\textwidth]{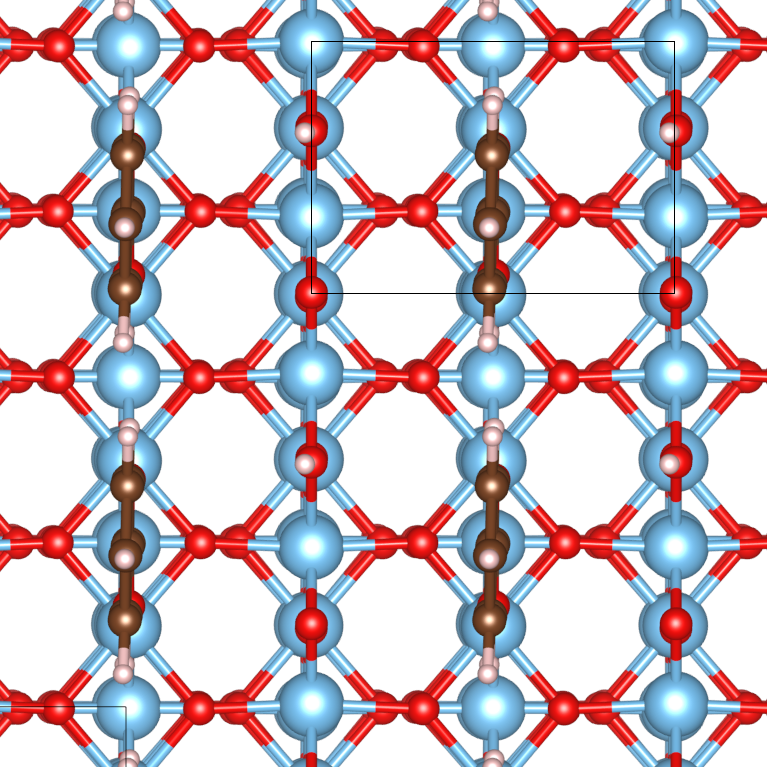}
\caption{The views of the GGA-optimized BB$_{\parallel, \text{ i.p.}}$ mode of adsorption of benzoic acid on the rutile $(110)$ surface, through the $[001]$, $[1\bar{1}0]$ and the ${110}$ directions. \label{ParaInGGA}}
\end{figure}

\begin{figure}[!hbtp]
\centering
\includegraphics[width = 0.30\textwidth]{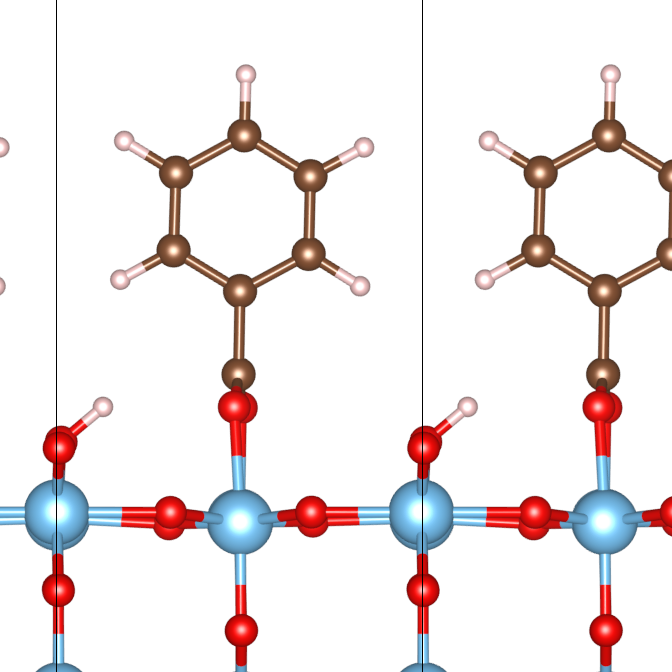}
\includegraphics[width = 0.30\textwidth]{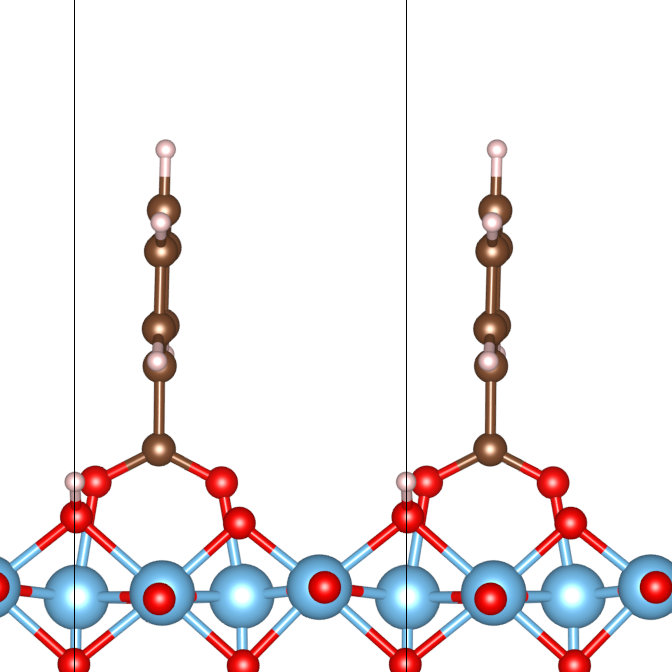}
\includegraphics[width = 0.30\textwidth]{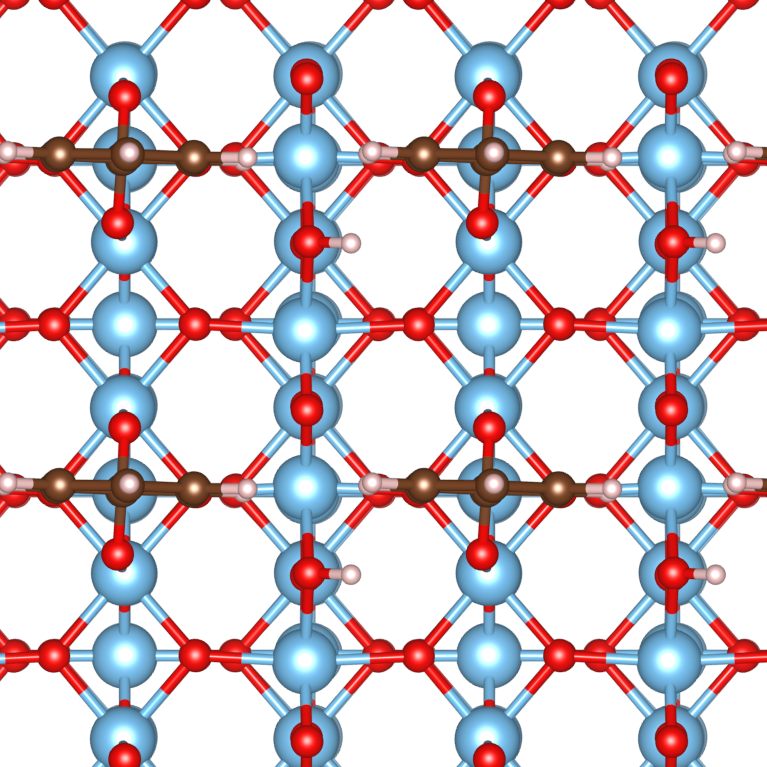}
\caption{The views of the GGA-optimized BB$_{\perp, \text{ i.p.}}$ mode of adsorption of benzoic acid on the rutile $(110)$ surface, through the $[001]$, $[1\bar{1}0]$ and the ${110}$ directions. \label{PerpInGGA}}
\end{figure}

\begin{figure}[!hbtp]
\centering
\includegraphics[width = 0.30\textwidth]{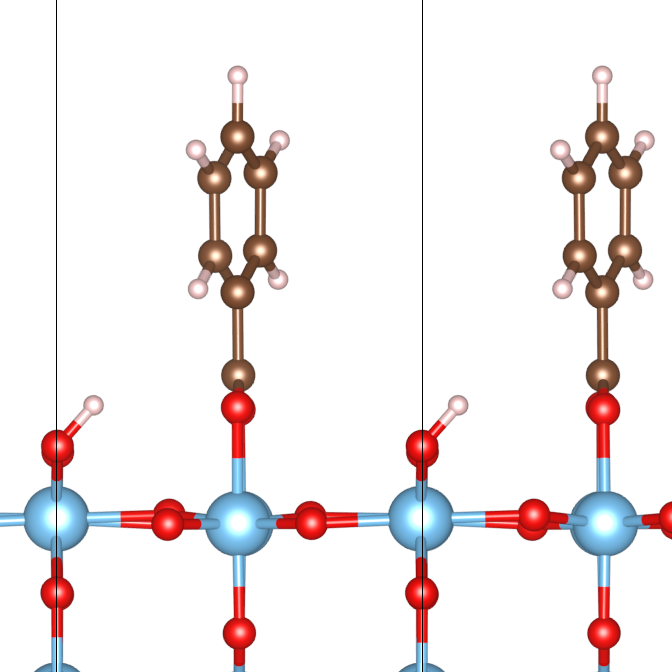}
\includegraphics[width = 0.30\textwidth]{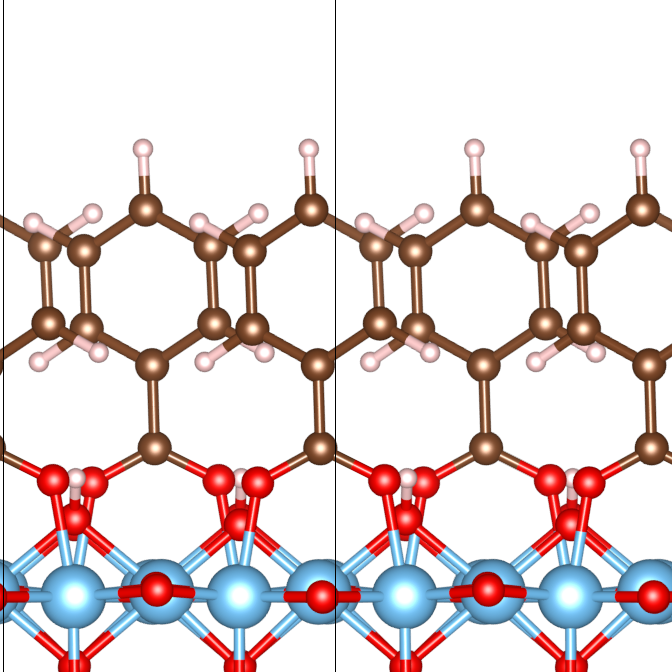}
\includegraphics[width = 0.30\textwidth]{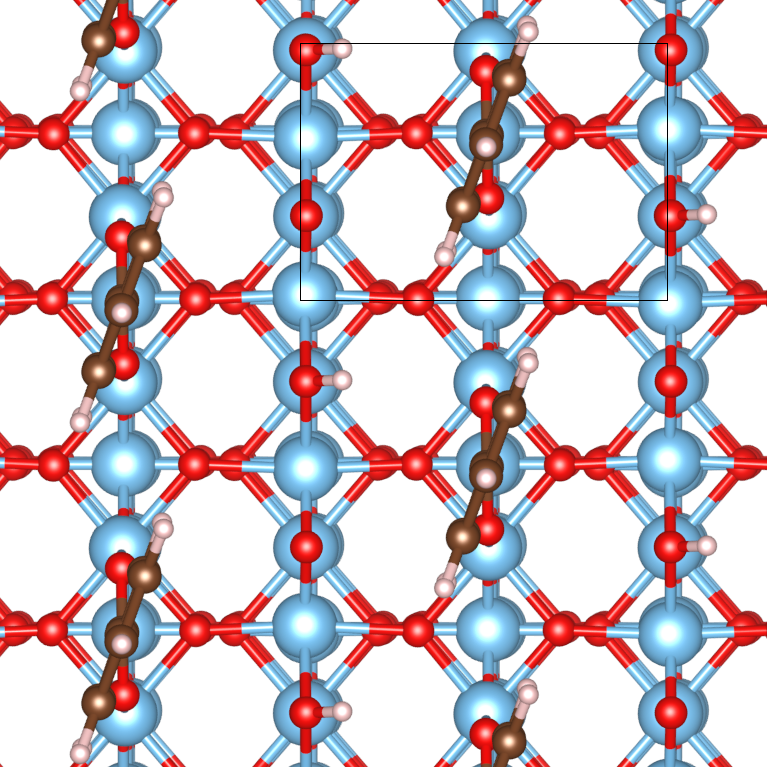}
\caption{The views of the GGA-optimized BB$_{\parallel, \text{ o.p.}}$ mode of adsorption of benzoic acid on the rutile $(110)$ surface, through the $[001]$, $[1\bar{1}0]$ and the ${110}$ directions. \label{ParaOutGGA}}
\end{figure}

\begin{figure}[!hbtp]
\centering
\includegraphics[width = 0.30\textwidth]{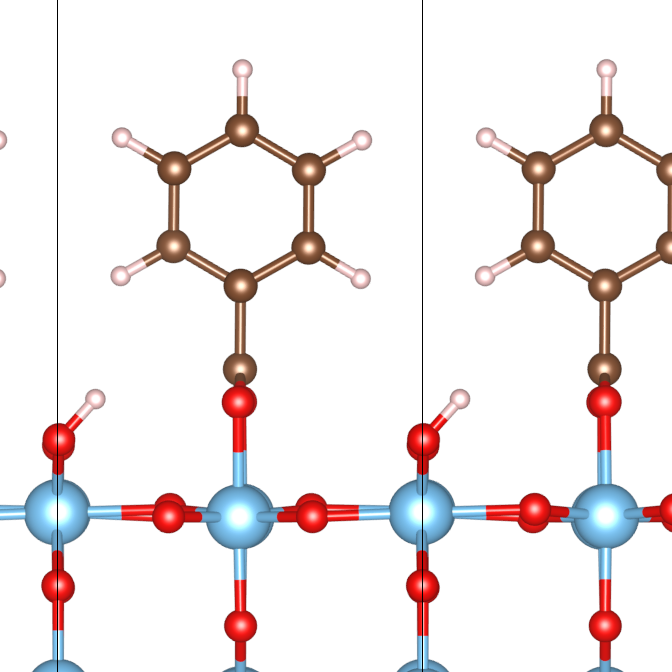}
\includegraphics[width = 0.30\textwidth]{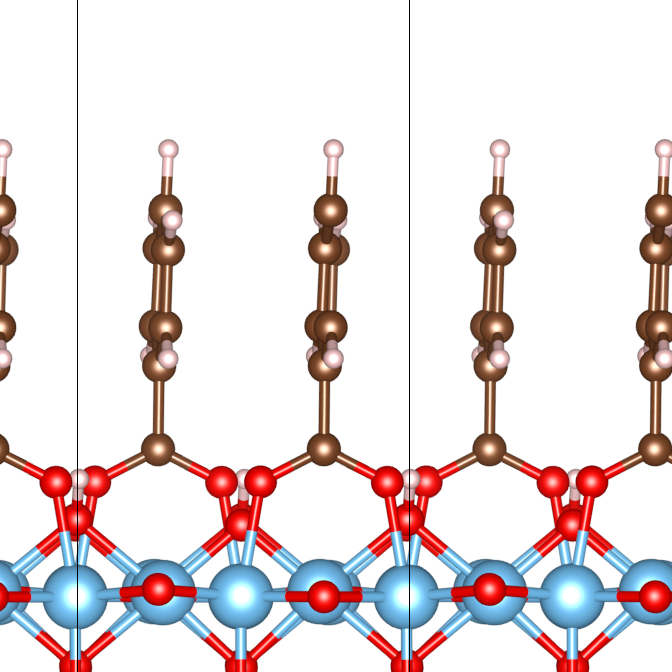}
\includegraphics[width = 0.30\textwidth]{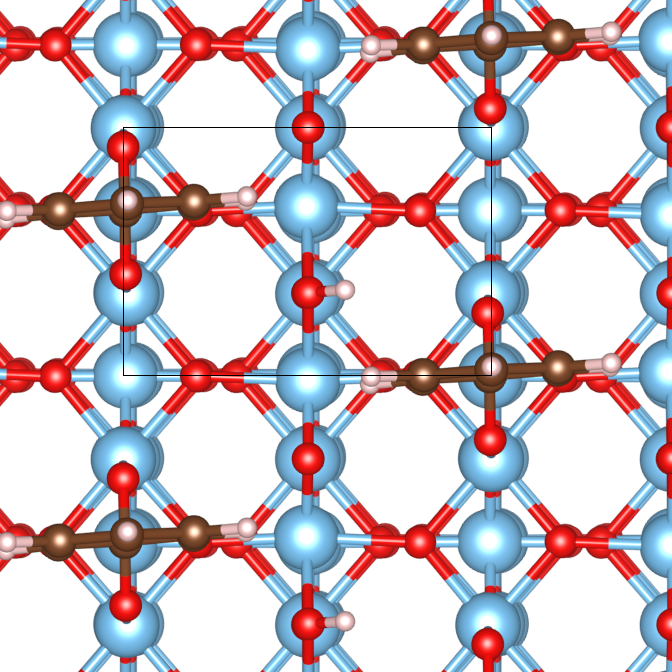}
\caption{The views of the GGA-optimized BB$_{\perp, \text{ o.p.}}$ mode of adsorption of benzoic acid on the rutile $(110)$ surface, through the $[001]$, $[1\bar{1}0]$ and the ${110}$ directions. \label{PerpOutGGA}}
\end{figure}

In terms of physical structures, all the four modes of adsorption, save the BB$_{\parallel, \text{ o.p.}}$ mode, exhibit adsorption structures close to those following the STM studies \cite{BenzRutNew}. Notable differences exist between the GGA-optimized adsorption configurations for the BB$_{\parallel}$ modes, and the idealized structures proposed to account for the STM observations. In both of the BB$_{\parallel}$ modes, buckling of the phenyl rings on the $[001]$ plane was observed. In the BB$_{\parallel, \text{ o.p.}}$ mode in particular, a rotation of the phenyl ring around the $[001]$ axis was observed. This could have come about as a result of the closer distances between the neighbouring phenyl rings of the adjacent benzoates along the $[001]$ rows, resulting in steric repulsions between the phenyl H atoms. In the BB$_{\perp}$ modes, such deviations from the idealized structures as proposed to model the STM studies \cite{BenzRutNew}, with the phenyl rings almost exactly aligned along the $[1\bar{1}0]$ direction. This can be due to the fact that the phenyl rings of the neighbouring benzoates along the $[1\bar{1}0]$ direction are further apart.

One particular interesting observation was the tilting of the H atoms in the cases of  BB$_{\parallel, \text{ o.p.}}$,  BB$_{\perp, \text{ i.p.}}$ and the BB$_{\perp, \text{ o.p.}}$ modes, towards the carboxyl O atoms of the respective modes, which were not observed in the LDA calculations. This is possibly due to the hydrogen bonding between the deprotonated H, and the carboxyl O. Such significant tilting, however, was not observed for the BB$_{\parallel, \text{ i.p.}}$ mode. The tilting observed in the former three modes, could be attributed towards the hydrogen bonding interactions between the deprotonated H, and the carboxyl O atoms. This however raises the question of why the same hydrogen bonding could not produce the same tilt on the H atom. This is possibly due to the fact that in the BB$_{\parallel, \text{ i.p.}}$ mode, the almost perfectly $[001]$-aligned phenyl rings, sterically hinders such an interaction through interactions between the rings' own H atoms and the carboxyl oxygens.

When the dissociated hydrogens are stripped from the surfaces, little structural changes were observed in all the modes mentioned. This was also true when Van der Waals' forces were taken into account when DFT-D2 was turned on, as the phenyl rings were observed to retain their orientations once the calculations attained relaxation. These thereby suggest that neither the presence or absence of H on the O$_{2\text{c}}$ sites had had significant effects on the physical orientations of the phenyl rings.

In terms of the adsorption energetics, the GGA-based calculations have shown that the adsorption energies have decreased from those of over ~2~eV/benzoate, to values much closer to 1.5~eV\cite{FormicRut110H}, \cite{FormicRut110AbInit}, \cite{FormAceRut110I} and \cite{FormAceRut110II}. In terms of the comparative energetic stabilities of the different adsorption modes, the GGA calculations have revealed that the BB$_{\parallel}$ modes were more energetically stable, than the BB$_{\perp}$ modes, in all the three pairs of modes. This is in clear contrast to the LDA results, where the BB$_{\perp}$ modes were the more energetically stable configurations out of all the three pairs. Part of the reason why in GGA the BB$_{\parallel}$ modes were more energetically stable, lies in the fact that now the inter-benzene ring distance between neighbouring rings along the $[110]$ direction was increased, resulting in less repulsion between them.

For the case of BB$_{\text{i.p.}}$ modes, the results agree with the conclusions drawn from STM, that the BB$_{\parallel, \text{i.p.}}$ mode is the more commonly observed mode in the $(1 \times 1)$-unreconstructed case. In fact, the energetic favourability was the greatest in this case, at a value of $~1.5$~eV, in contrast to the case of LDA, where the BB$_{\perp, \text{i.p.}}$ mode was calculated to be much more energetically stable (by ~1.2~eV). The comparative greater energetic stability in the  BB$_{\parallel, \text{i.p.}}$ mode can now be described in terms of greater spacing between the neightbouring benzene rings, being increased by nearly $0.1$~~\AA, resulting in less sideways repulsion between the sidemost H atoms.

In the other pairs of BB$_{\parallel}$ and BB$_{\perp}$ modes, although the adsorption energy values were closer to the above range, they did not agree with the models proposed in the experimental studies, in which both the BB$_{\text{o.p.}}$ and the BB$_{1 \times 2}$ modes had the BB$_{\perp}$ modes were described to be the predominant form observed in STM. This is however explainable by the presence of co-adsorbed H on the O$_{2\text{c}}$ on the $[001]$ ridges, which serve to repel H atoms on the rotated benzene rings.

\begin{figure}[!hbtp]
\centering
\includegraphics[width = 0.60\textwidth]{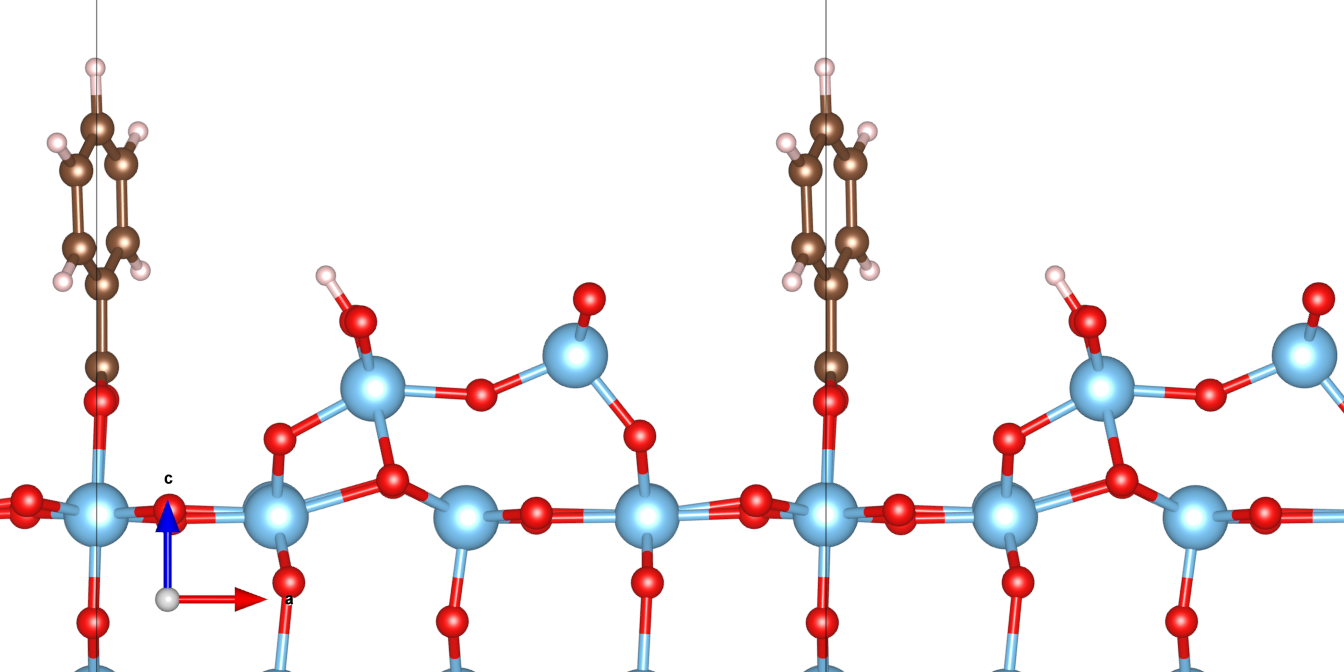}
\includegraphics[width = 0.30\textwidth]{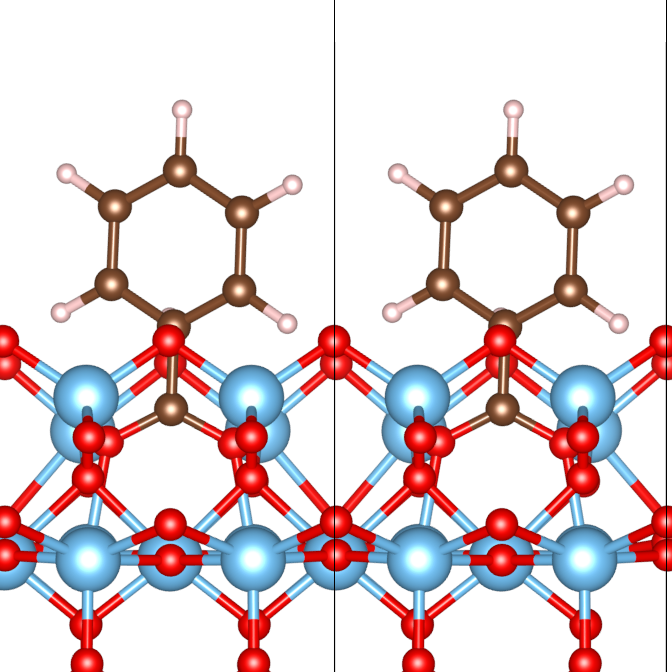}
\includegraphics[width = 0.60\textwidth]{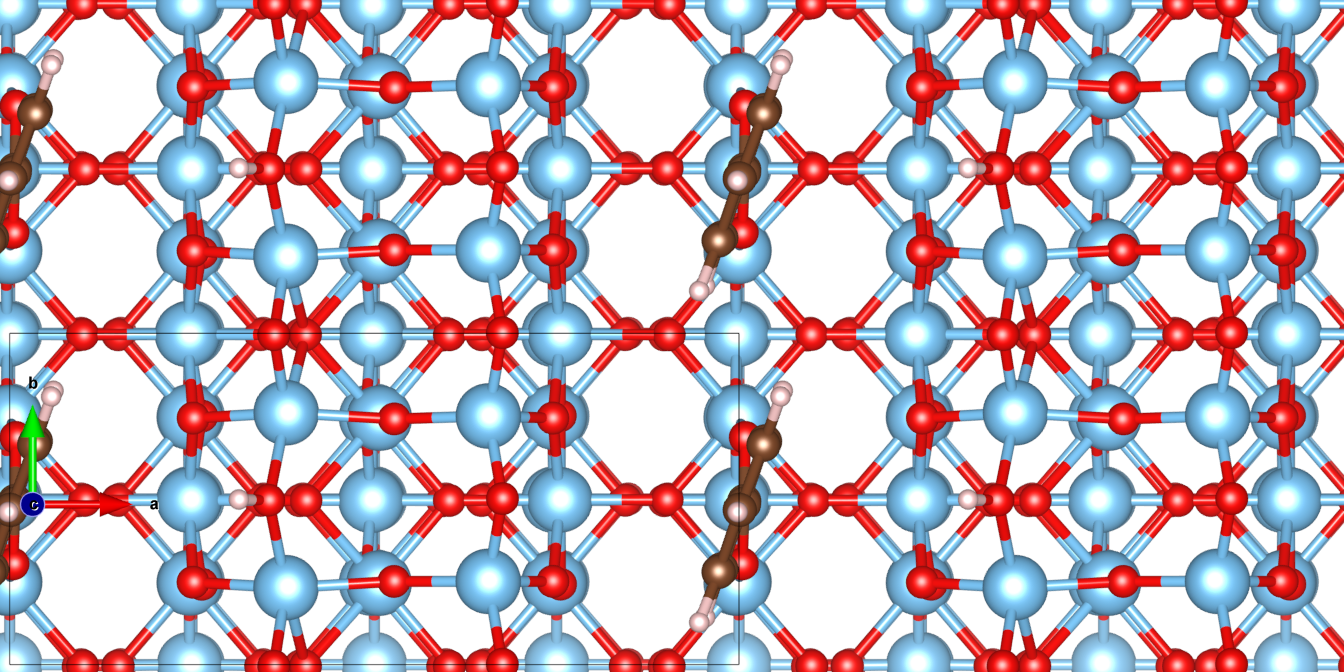}
\caption{The views of the GGA-optimized BB$_{\parallel, \text{ i.p.}}$ mode of adsorption of benzoic acid on the rutile $(110)-(1 \times 2)$ reconstructed surface, through the $[001]$, $[1\bar{1}0]$ and the ${110}$ directions. \label{ParaReconGGA}}
\end{figure}

\begin{figure}[!hbtp]
\centering
\includegraphics[width = 0.60\textwidth]{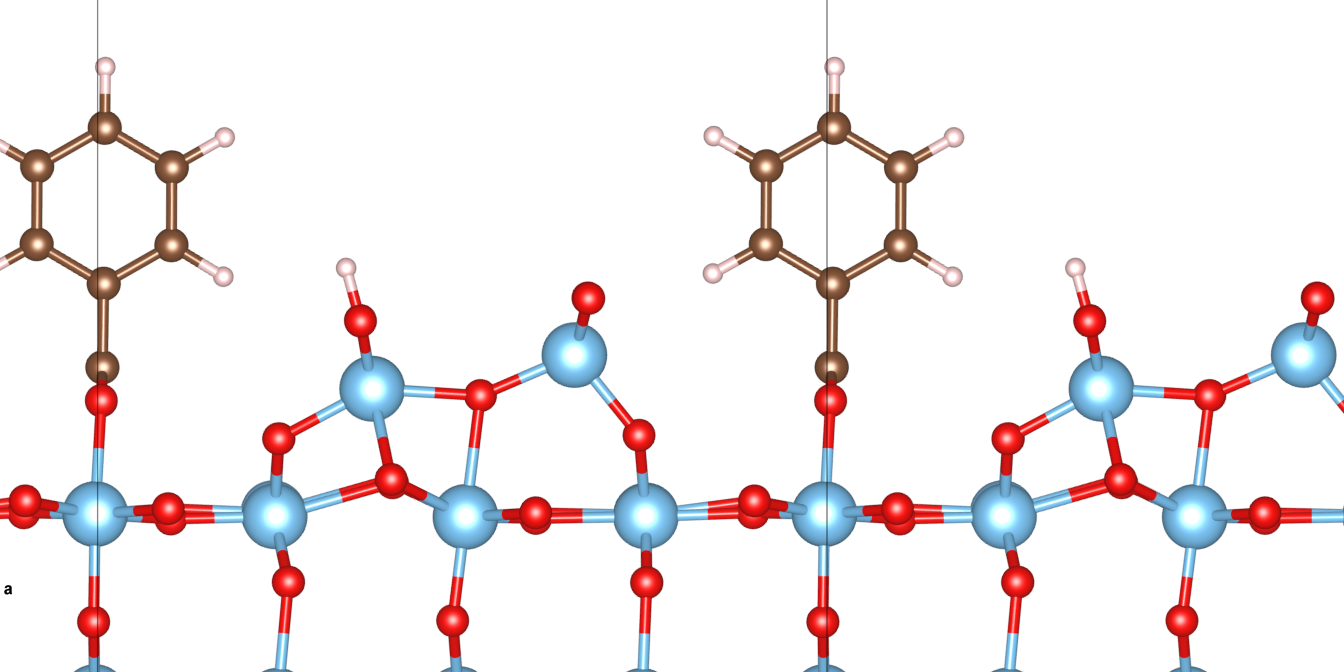}
\includegraphics[width = 0.30\textwidth]{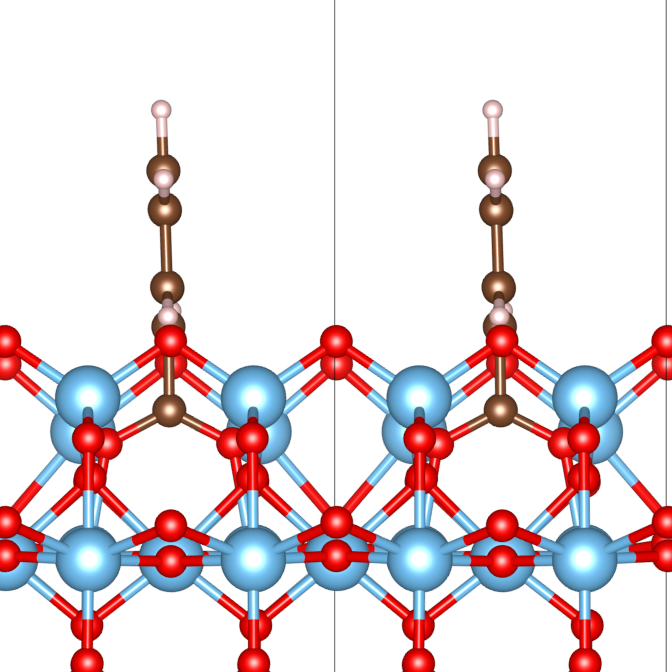}
\includegraphics[width = 0.60\textwidth]{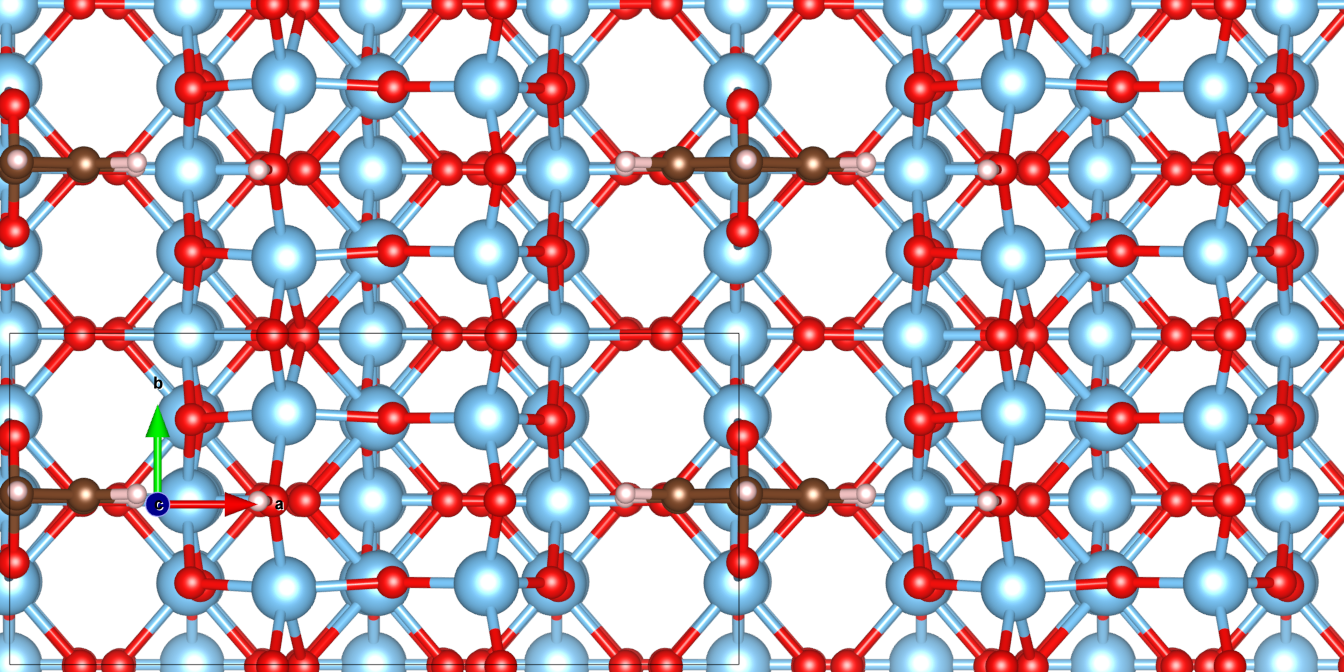}
\caption{The views of the GGA-optimized BB$_{\perp, \text{ i.p.}}$ mode of adsorption of benzoic acid on the rutile $(110)-(1 \times 2)$ reconstructed surface, through the $[001]$, $[1\bar{1}0]$ and the ${110}$ directions. \label{PerpReconGGA}}
\end{figure}

When the acid was dissociatively adsorbed onto the rutile $(110)$ in the manners of BB$_{\text{o. p.}}$ modes, similar to the results produced in LDA, the difference in terms of the energetic stabilities of this pair of BB$_{\parallel}$ and BB$_{\perp}$ modes was the narrowest of the three at $~0.25$~eV, though with the BB$_{\parallel, \text{o. p.}}$ mode being more energetically stable this time. This gap is also larger than that reported for the corresponding pair in the LDA calculations, where the BB$_{\perp, \text{o. p.}}$ mode was more energetically stable by just $~0.05$~eV/molecule. The GGA results suggest that rotation of the benzene rings by $90^{\circ}$ in order to avoid the steric repulsion between the rings along the $[001]$ directions, as well as the weak H interactions between the benzyl H and the O$_{2\text{c}}$ of the $[001]$ rows, do not energetically stabilize the structure, and that rotation of benzene rings themselves can serve as an energetically destabilizing factor.

For the adsorption on the reconstructed surface, while the BB$_{\parallel, 1 \times 2}$ mode was calculated to be the more energetically stable, as compared to the BB$_{\text{i.p.}}$ modes, the energetic favourability was much lower at $~0.25$~eV/molecule. Furthermore, as compared to the corresponding pair of BB$_{1 \times 2}$ modes relaxed in the LDA calculations, which showed that the BB$_{\perp, 1 \times 2}$ mode was more energetically stable by $~0.9$~eV/molecule, a much greater difference in the favour of the other adsorption pattern. Again, this suggests that rotation of the benzene ring in itself might have been an energetically destablizing effect on the entire set up.

When the co-adsorbed hydrogens were removed, the structures changed very little, with the most significant being the partial restoration of the symmetry of Ti$_{2}$O$_{3}$ reconstructed ridges. This shows that the presence of H atoms on the O$_{2\text{c}}$ ridges does not affect hydrogen bonding interactions between most parts of the surface and any part of the benzoate. In terms of ionic adsorption energetics themselves, the comparative energetic stability of the BB$_{\parallel}$ over the corresponding BB$_{\perp}$ mode was confirmed, while in all of the cases, adsorption of benzoate ions without the co-adsorption of the dissociated H had a greater energetically stabilizing effect than dissociative co-adsorption of the both ions on the same surface with the same geometry.

One interesting thing to note is that while for the BB$_{\text{i. p.}}$ and BB$_{\text{o. p.}}$ modes, the E$_{\text{ads}}$(\textbf{no H}) values are more , while that for the BB$_{1 \times 2}$ modes the E$_{\text{ads}}$(\textbf{no H}) values appeared to be far larger than the corresponding E$_{\text{ads}}$ values, over twice values of the E$_{\text{ads}}$ obtained for dissociative co-adsorption. As discussed earlier, a partial restoration of symmetry of the Ti$_{2}$O$_{3}$ reconstructed ridges was observed following the removal of the dissociated H, and this perhaps provided the additional energetic stabilization. However, due to the large increase in the values of E$_{ads}$ involved, and the values for E$_{\text{ads}}$(\textbf{no H}) differed little regardless of simulation cell used and the same E$_{\text{slab}}$ values were used for both the dissociative co-adsorptive and the ionic adsorptive cases, further investigations shall be needed to determine the exact cause of this unusually large increase in the adsorption energy when the dissociated hydrogens are not taken into account.

\subsubsection{GGA-DFTD2 Calculations}
Although Van der Waals' forces are weak in comparison to hydrogen bonding when comes to inter-ionic interactions between adsorbates, they can come into play when large non-polar components come close to each other, and especially so in the case of phenyl rings with extensive $\pi$-orbitals being close with each other, as seen in Figures \ref{ParaInGGA}-\ref{PerpReconGGA}. Indeed, taking into account of Van der Waals' forces through GGA+DFTD2 calculations has improved the energetic stabilities for all six different modes, each being at least ~0.7 eV/molecule more stable after implementation of GGA-DFTD2 instead of just GGA (see Table \ref{DFTD2Ads}). In spite of these changes, the BB$_{\parallel}$ modes are still more energetically stable compared to their BB$_{\perp}$ counterparts, when the co-adsorbed hydrogens are taken into account.

When just considering the case of ionic adsorption, the changes in E$_{\text{ads}}$ values become even more pronounced, and especially so in the case of benzoates' adsorptions on the $(1 \times 2)$-reconstructed surface. Not only are the E$_{\text{ads}}$ values abnormally large, but also that the BB$_{\perp, 1\times2}$ mode has now become significantly more energetically stable as compared to the  BB$_{\parallel, 1\times2}$ mode (-7.87 vs -5.14~eV/benzoate). These anomalies shall be discussed in further detail in the \''Discussions\'' section.

\begin{table} 
\begin{tabular}{|p{3.5cm}|p{3.5cm}|p{3.5cm}|p{3.5cm}|}  
\hline
\textbf{Adsorption Mode} & E$_{\text{ads}}$(\textbf{dis}) & E$_{\text{ads}}$(\textbf{no H}) \\
\hline
BB$_{\parallel, \text{ i.p.}}$ (Figure \ref{ParaInGGA}) & -3.55 & -3.95 \\
\hline
BB$_{\perp, \text{ i.p.}}$ (Figure \ref{PerpInGGA}) & -2.12 & -2.52 \\
\hline
BB$_{\parallel, \text{ o.p.}}$ (Figure \ref{ParaOutGGA}) & -2.48 & -2.78 \\
\hline
BB$_{\perp, \text{ o.p.}}$ (Figure \ref{PerpOutGGA}) & -2.23 & -2.55 \\
\hline
BB$_{\parallel, \, 1\times2}$ (Figure \ref{ParaReconGGA}) & -2.04 & -5.14* \\
\hline
BB$_{\perp, \, 1\times2}$ (Figure \ref{PerpReconGGA}) & -1.72 & -7.87* \\
\hline
\end{tabular} 
\caption{Adsorption energies for ionic adsorptions of benzoates on the rutile $(110)$ surface, in~eV/\AA, calculated using GGA+DFTD2, with ''*'' representing anomalies in values. \label{DFTD2Ads}}
\end{table}

\subsection{Simulated STM Results}
\begin{figure}[!hbtp]
\centering
\includegraphics[width = 0.45\textwidth]{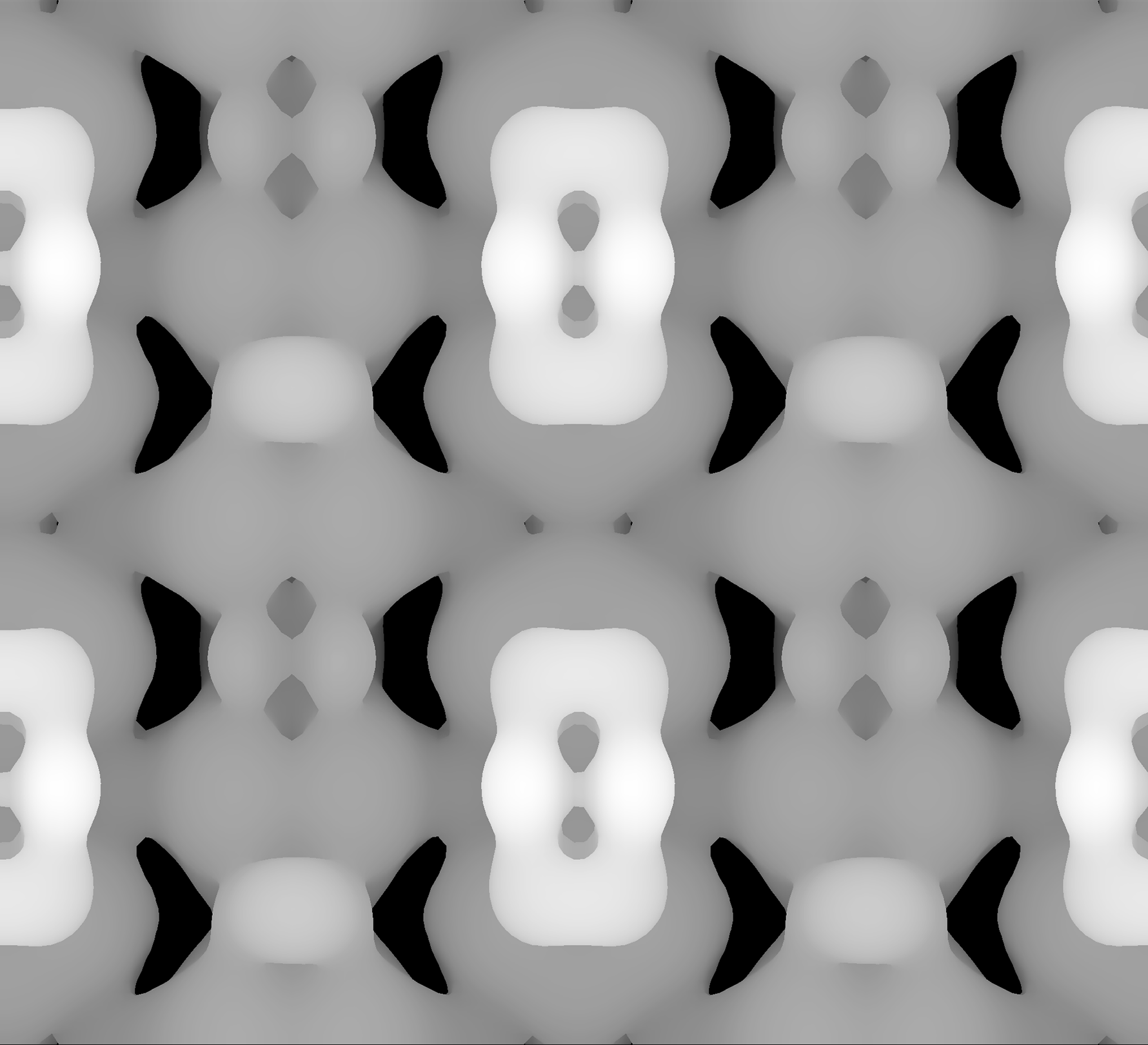}
\includegraphics[width = 0.45\textwidth]{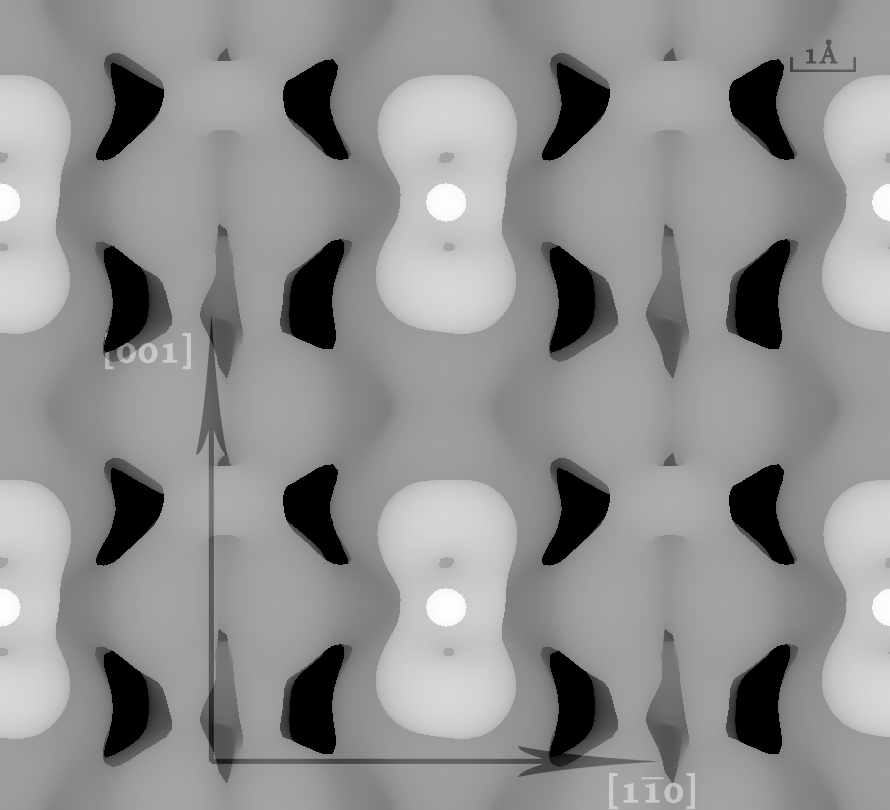}
\includegraphics[width = 0.45\textwidth]{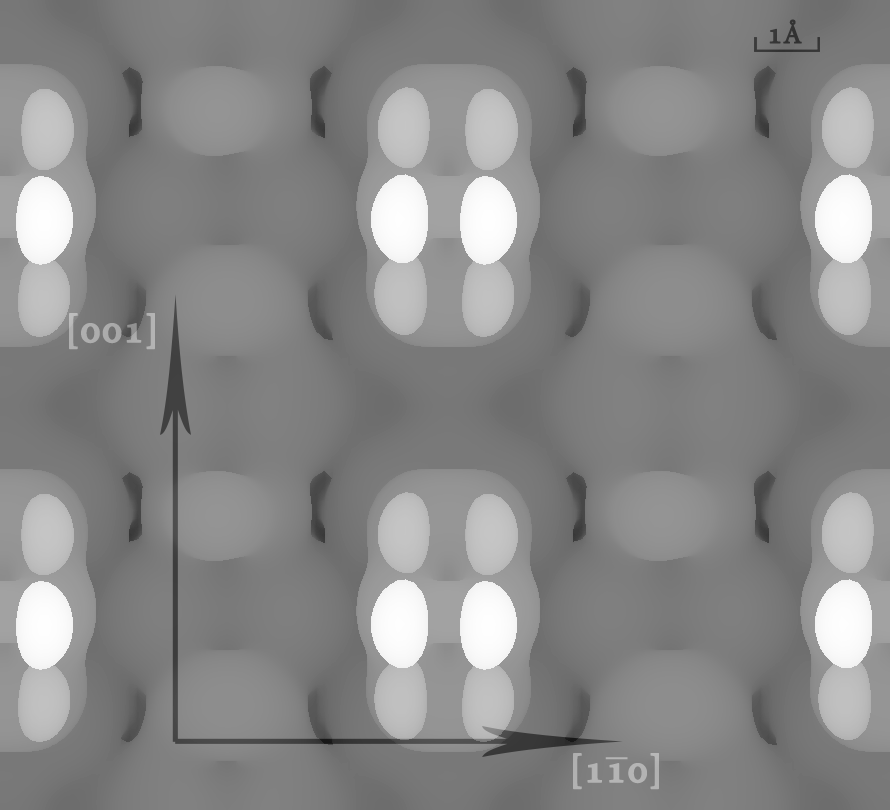}
\includegraphics[width = 0.45\textwidth]{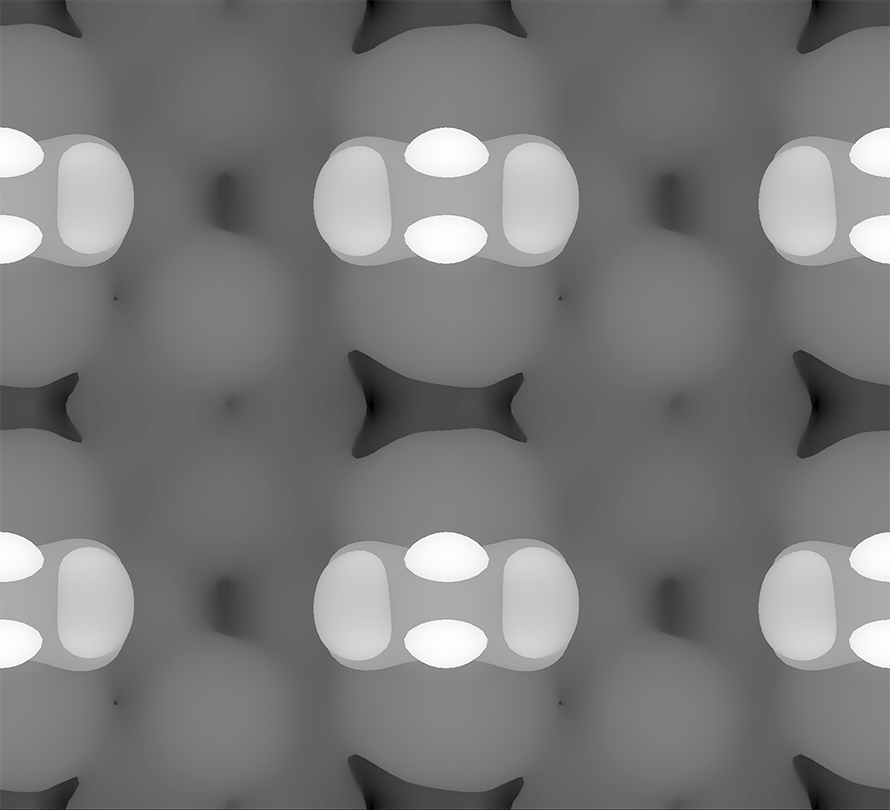}
\caption{Simulated STM images for the BB$_{\parallel, \text{i. p.}}$ (left) and BB$_{\perp, \text{i. p.}}$ (right) modes, at high (top) and low (bottom) charge valued isosurfaces, i.e. high and low tunneling current, revealing the positions of the benzoates (top) and the orientations of the benzene rings (bottom) respectively. \label{BBipSTM}}
\end{figure}

\begin{figure}[!hbtp]
\centering
\includegraphics[width = 0.45\textwidth]{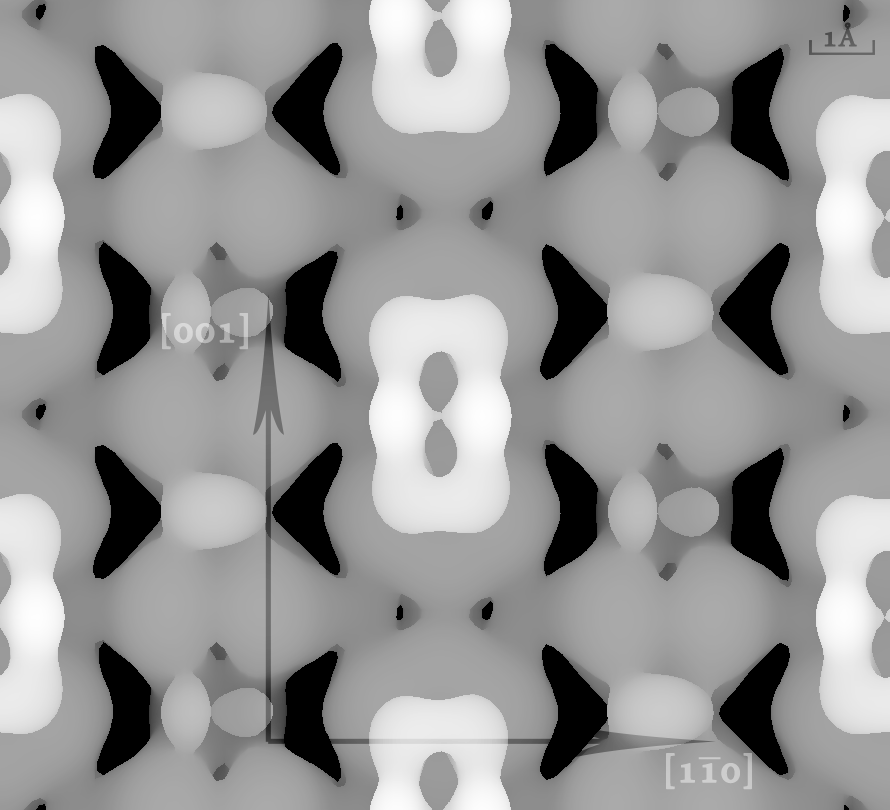}
\includegraphics[width = 0.45\textwidth]{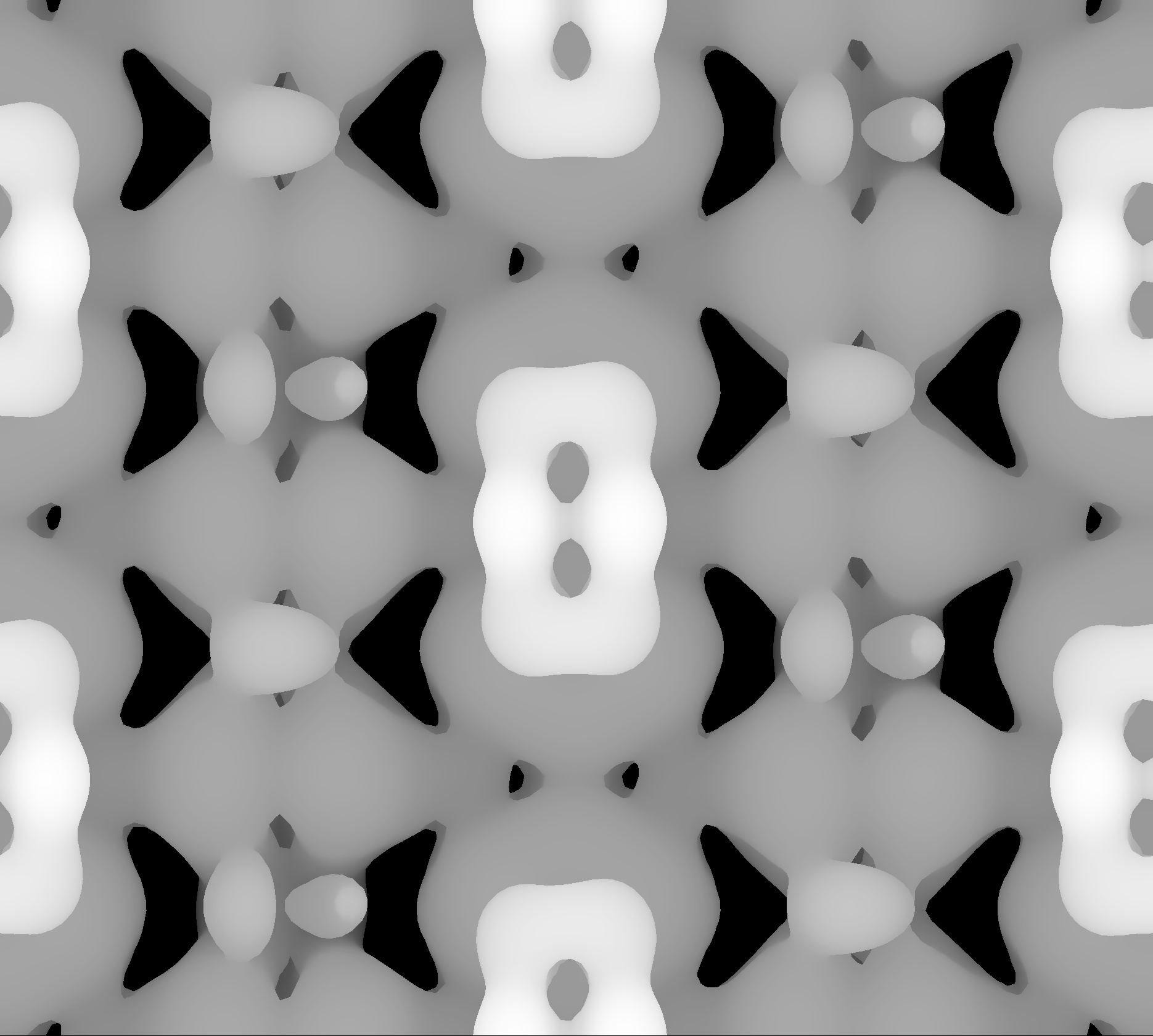}
\includegraphics[width = 0.45\textwidth]{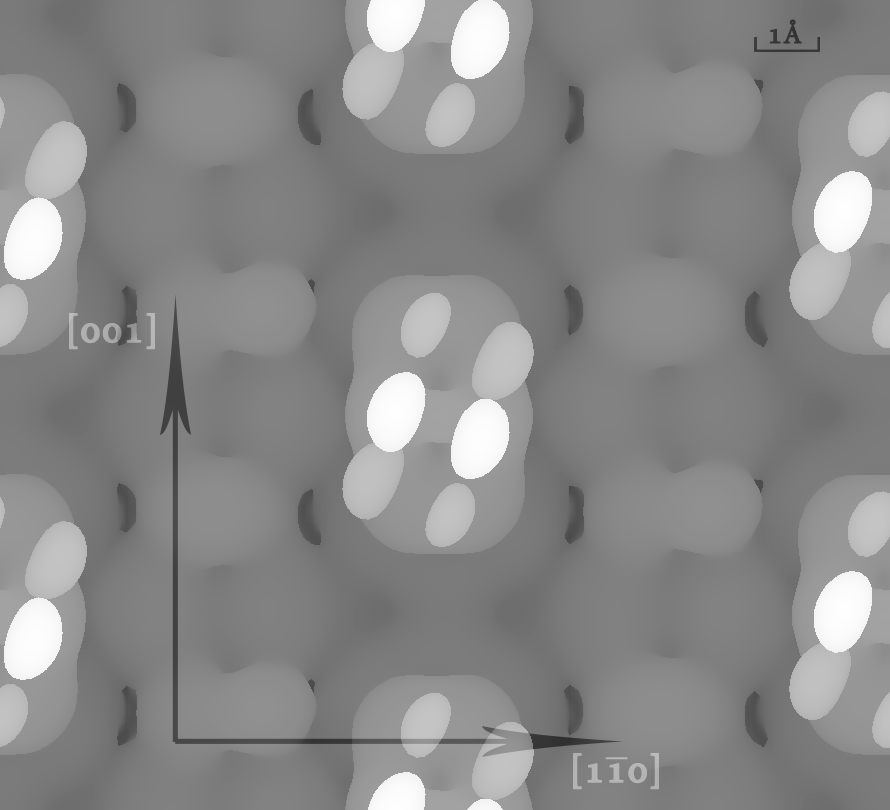}
\includegraphics[width = 0.45\textwidth]{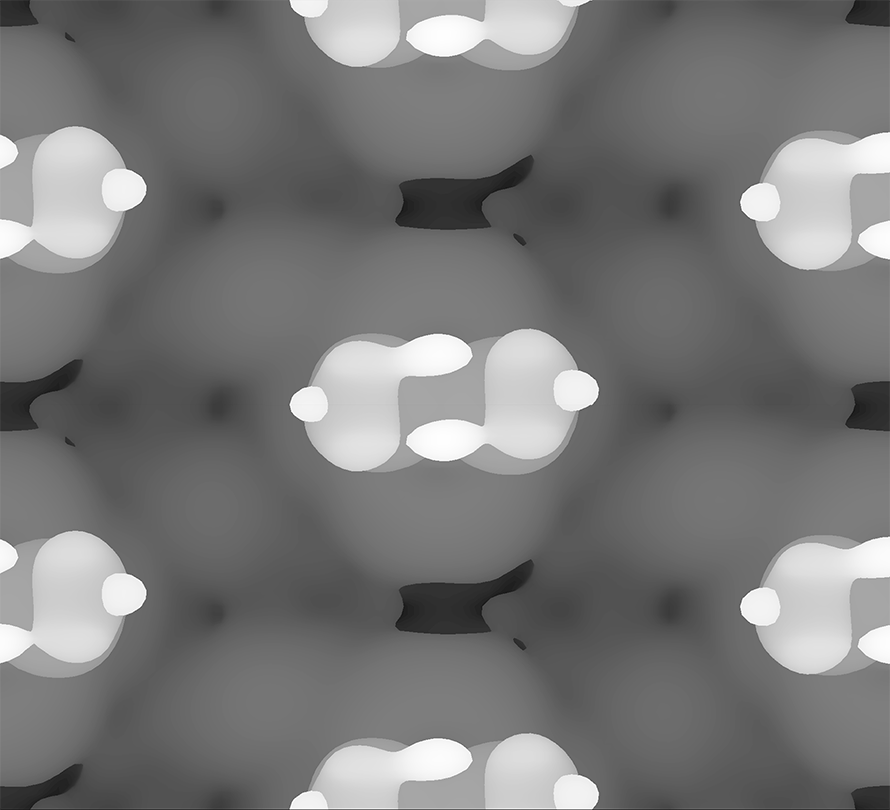}
\caption{Simulated STM images for the BB$_{\parallel, \text{o. p.}}$ (left) and BB$_{\perp, \text{o. p.}}$ (right) modes, at high (top) and low (bottom) charge valued isosurfaces, revealing the positions of the benzoates (top) and the orientations of the benzene rings (bottom) respectively. \label{BBopSTM}}
\end{figure}

In the experimental STM studies, the STM images produced presented the sites of benzoate adsorption as bright spots, elongated along the directions of the alignments of the phenyl rings\cite{BenzRutNew}, with no direct visual evidence of co-adsorbed hydrogens, except for low coverages on the $(1 \times 2)$-reconstructed surface. The simulated STM images produced have replicated the same arrays of bright spots representing the two-dimensional periodities of the benzoate adsorbates (see Figures \ref{BBipSTM} and \ref{BBopSTM}), while the rutile $(110)$ surface itself has been shown to be dark in comparison to the adsorption sites.

In the limit representing high charge isosurfaces i.e. large currents, the positions and the shapes of the bright regions representing the benzoate adsorption sites were mainly determined by the positioning of the carboxylate groups, rather than the orientations of the benzene rings. The latter only started to influence the appearances of bright regions when the charge isosurfaces are set at values sufficiently low (i.e. limits of low currents), such that the rutile $(110)$ substrate surface itself loses its distinctive visual features, such as distinct dark regions as shown in Figure \ref{BBipSTM}(top). Where the benzene rings were imaged in the limit of low currents, they appeared either as double rows of trios of bright spots, with the brightest at the centre representing the topmost C atom of the ring in the case of the benzene ring being aligned with the $[001]$ direction; or as twin central bright spots accompanied by darker, kidney-shaped spots on either side. It is also observed while in the limit of high currents where the carboxylate groups were distinctively imaged, positions of the co-adsorbed H can be seen from the differing orbital shapes in the $[001]$ O$_{2\text{c}}$ rows, with circular-shaped dim spots representing the 1s orbitals of H and dumbell-shaped ones representing the 2p orbitals of the O$_{2\text{c}}$ atoms.

In the BB$_{\text{i. p.}}$ modes, whose simulated STM images are as shown in Figure \ref{BBipSTM}, revealed orientations of rectangular rows of bright spots, with their widths aligned along the planes of the benzene rings determined through DFT calculations. The patterns of the bright spots representing the benzene rings however differed between the BB$_{\parallel, \text{i. p.}}$ and the BB$_{\perp, \text{i. p.}}$ modes, with the former as parallel trios of bright spots along the $[001]$-direction, and the latter as two central bright spots sandwiched by two larger merged dimmer spots aligned along the $[1\bar{1}0]$-direction. 

In the BB$_{\text{o. p.}}$ modes, the simulated STM images generated are as shown in Figure \ref{BBopSTM}. The bright spots representing the $(2 \times 2)$ overlayer symmetry can be observed in both cases, in agreement with STM results obtained from experimental studies, where such features were observed in some regions alongside the bright dots representing the $(1 \times 1)$ symmetry on the unrecontructed surface. The orientations of the bright spots again reflected the orientations of the benzene rings as shown in Figures \ref{ParaOutGGA} and \ref{PerpOutGGA}. However, whereas the benzene ring was imaged as duo rows of bright spots in the BB$_{\parallel, \text{i. p.}}$ mode, in the BB$_{\perp, \text{i. p.}}$ mode the darker spots around the central bright spots merged, forming kidney-shaped spots.

\begin{figure}[!hbtp]
\centering
\includegraphics[width = 0.45\textwidth]{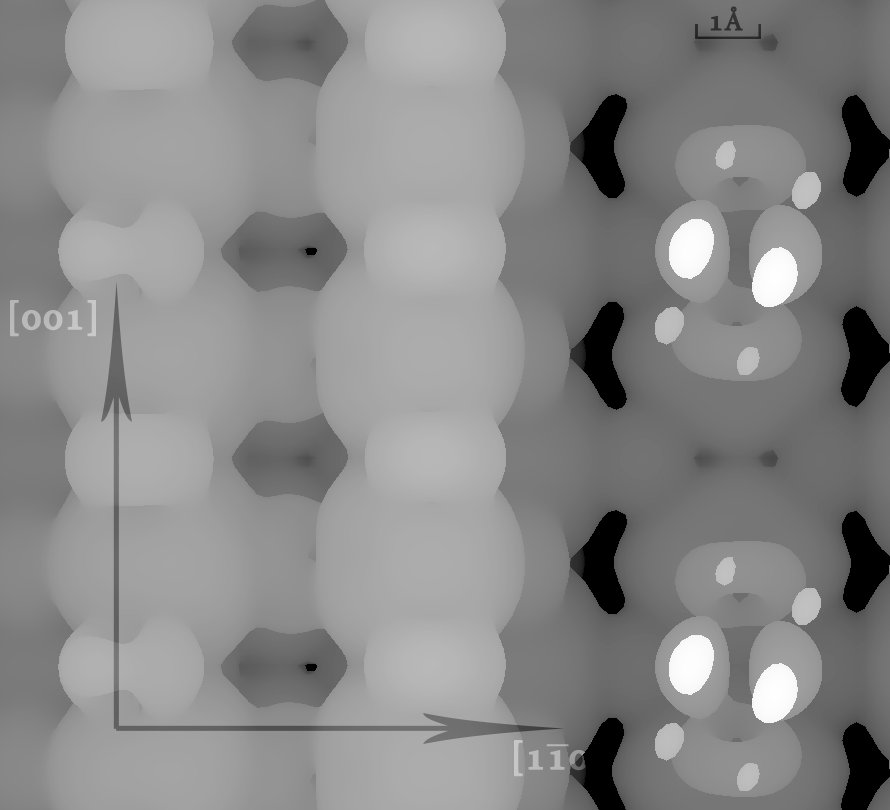}
\includegraphics[width = 0.45\textwidth]{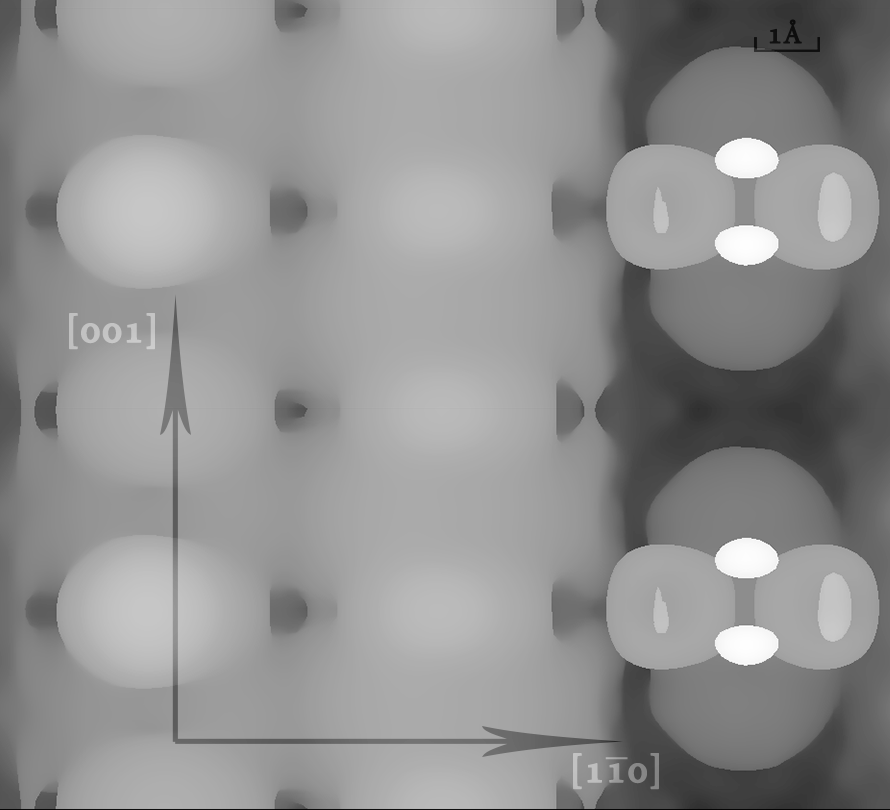}
\caption{Simulated STM images for the BB$_{\parallel, 1 \times 2}$ (left) and BB$_{\perp, 1 \times 2}$ (right) modes, in the limit of low charge isosurfaces, revealing the orientations of the benzene rings. \label{BBReconSTM}}
\end{figure}

For the case of BB$_{1 \times 2}$ modes, in both cases the reconstructed Ti$_{2}$O$_{3}$ ridge appeared as large bright bands along the $[001]$ directions, with features showing the ridges being largely indistinguishable. Again, although the double rows of three bright spots were observed when the ring was aligned along the $[001]$ direction, the dimmer pairs of spots on the sides of the central bright spots became merged in the case of the benzene ring being aligned with the $[1\bar{1}0]$ direction.

\section{Discussions}
For adsorption energetics, GGA and GGA+DFTD2 simulations have shown the BB$_{\parallel, \text{i. p.}}$ to be the most energetically stable out of all the BB modes of adsorptions on the unreconstructed surface, when co-adsorbed with the dissociated hydrogens. Being the most observed pattern in the experimental STM studies of the unreconstructed surface, this is an expected result. In terms of the simulations themselves, this shows that when hydrogen bonding and/or Van der Waals' forces are taken into account, the BB$_{\parallel, \text{i. p.}}$ mode is the most stable.

Examining the three pairs of BB$_{\parallel}$ and BB$_{\perp}$ modes of adsorption, in GGA and GGA+DFTD2-based calculations, it is clear that on the non-reconstructed surface, simulations have shown that for both $(1 \times 2)$- and $(2 \times 2)$-symmetric patterns, the BB$_{\parallel}$ modes were shown to be more energetically stable in both cases. These observations, in both the experimental and computational DFT studies, can be accounted for by the long distances between the O$_{2\text{c}}$ atoms and the nearest phenyl H atoms when the benzene ring is rotated, making hydrogen bonding between the two too weak to be a significant stablizing factor.

In terms of the $(2 \times 2)$-symmetric patterns on the rutile $(110)$ surface, the BB$_{\parallel, \text{o. p.}}$ modes were shown to be more energetically stable through GGA and GGA+DFTD2 calculations. This again is in line with the experimental observations, which revealed alignment of the benzene rings with the $[001]$ direction. The proposed reasons for such observations were again that the nearest distances between the O$_{2\text{c}}$ atoms and the phenyl H atoms being too far apart to make hydrogen bonding a significant stablizing force. In comparison the BB$_{\parallel, \text{i. p.}}$ mode, the corresponding  BB$_{\parallel, \text{o. p.}}$ mode appeared less energetically stable (see Figures \ref{DissAds} and \ref{IonAds}). This again agrees with the findings of experimental STM studies, where such patterns were comparatively rare relative to the $(2 \times 1)$-symmetric patterns.

In the case of the $(1 \times 2)$-reconstructed rutile $(110)$ surface, however, the results of the DFT calculations contradict the conclusions reached in the STM studies. In both the cases of GGA and GGA+DFTD2 calculations, it was the BB$_{\parallel, 1 \times 2}$ configuration that was more energetically stable in all the three pairs of simulations, instead of the $[1\bar{1}0]$-aligned benzene rings as observed in the experimental STM images. The likely explanation for this were the asymmetric distortions in the Ti$_{2}$O$_{3}$ reconstruction ridges produced in the optimized structures, with the predicted energetic stablization effects from hydrogen bonding between the O$_{2\text{c}}$ atoms on the reconstructed ridges and the nearest phenyl H being less than predicted.

Perhaps the single greatest anomaly in the results lies in the abnormally large ionic adsorption energy values for benzoates on the $1 \times 2$-reconstructed surface - see Figure \ref{DFTD2Ads}. We found that, for the ionic adsorption on the reconstructed surface, the energies appeared to be anomalous. We include the values for completeness, but are unable to explain their large values, which we will continue to investigate.

In terms of the simulated STM images, the images generated revealed the bright spots corresponding to sites of benzoate adsorption, with the orientations of the rings much more clearly delineated than the experimental STM images do. The arrays of positions of the bright spots also corresponded to the 2D symmetries of the adsorption monolayers in the BB$_{\text{i. p.}}$,  BB$_{\text{o. p.}}$ and  BB$_{1 \times 2}$ configurations.

For both the BB$_{\text{i. p.}}$ and the BB$_{\text{o. p.}}$ modes, in the limit of sufficiently low valued charge isosurfaces such that the benzene rings start to be imaged distinctively, the co-adsorbed H atoms lose their distinguishability from their surroundings, whereas in the limit of isosurfaces of high charge values, the vacant O$_{2\text{c}}$ sites could be distinguished from the ones bound by dissociated H by the 2p orbitals imaged. The simulated STM images in the low charge value limits reveal features which are consistent with those observed in the experimental STM images produced of benzoates on the unreconstructed surface, most notably the elongations of the bright regions along the directions of alignments of the benzene rings, as well as the absence of features reflecting the co-adsorption of dissociated H. This suggests that perhaps even if STM images do not directly image the co-adsorbed H ions, it does not mean that such co-adsorption did not take place.

On the reconstructed surfaces, the simulated STM images revealed central bright bands representing the Ti$_{2}$O$_{3}$ reconstructed rows similar to those observed in the experimental STM images, and that no distinguishable features representing co-adsorption of dissociated H ions. This is however still in agreement with experimental results as co-adsorbed H ions were not imaged at high coverages. Again, the lack of direct visual evidence for co-adsorption of dissociated protons does not necessarily imply complete absence of it.

\section{Conclusions}
We have studied the different patterns for the bidentate bridging dissociative adsorption of benzoic acid on both the unreconstructed and $(1 \times 2)$-reconstructed surfaces through DFT methods LDA, GGA and GGA+DFTD2, as implemented through VASP, following both $(2 \times 1)$ and $(2 \times 2)$ symmetries. In all the three DFT methods studied, both with and without the dissociated H, the BB$_{\parallel, \text{i. p.}}$ pattern was invariably found to be the most energetically stable out all the six BB patterns of adsorption. These results support STM studies \cite{BenzRutNew} which revealed that on the unreconstructed $(110)$ surface, where the prevalent pattern was that of BB$_{\text{i. p.}}$, with the orientation of the benzene rings aligned along the $[001]$ direction.

In the case of $(1 \times 2)$-reconstructed surface however, the BB$_{\parallel, 1 \times 2}$ mode was found to be more stable, in contrast to the experimental STM studies' results\cite{BenzRutNew} where the benzene rings were observed to have rotated by 90$^{\circ}$, with hydrogen bonding between the phenyl hydrogens and the O$_{2\text{c}}$ atoms along the reconstructed ridges being the reason for the stablization of this orientation. 

In the case of simulated STM images, the same overlayer patterns of bright dots was reproduced for the proposed models, elongations of the bright spots representing the alignments of benzene rings were observed, which were not obvious in the case of experimental STM studies. These alignments were only deduced by line profiles taken along the $[001]$ and the $[110]$ directions in the experimental STM studies. In the different limits of high and low-valued charge isosurfaces, the simulated STM images have also revealed the carboxylate and the benzene ring groups respectively, while in experimental STM images, these structures were not clearly distinguished in the bright spots representing the sites of benzoate adsorption.


\section{Acknowledgement}
We thank Dr Geoff Thorton from the Chemistry Department of University College London for providing the model for the rutile $(110)$-$(1 \times 2)$-reconstructed surface, the Research Computing group of the Internet Services Division of University College London, and Jakub Vrtny, Conn O'Rourke and Umberto Terranova for valuable discussions. This research work is funded by UCKG Limited, China.

\end{document}